\documentstyle[11pt,aaspp4]{article}

\lefthead{Latter et al.}
\righthead{HST/NICMOS observations of NGC 7027}

\def\n7027{NGC 7027}
\def\h2{H$_2$}
\def\s1{$v = 1 \to 0$ S(1)}

\def\about{$\approx$}

\def\Paa{P$\alpha$}
\def\Pa{P$\alpha$}
\def\Brg{Br$\gamma$}

\def\etal{et al.\ }
\def\eg{e.g.,\ }
\def\ie{i.e.,\ }

\def\cm3{cm$^{-3}$}
\def\kcm3{{\rm cm}^3~{\rm s}^{-1}}
\def\fcm3{{\rm cm}^{-3}}
\def\kms{km ${\rm s}^{-1}$}
\def\msol{M$_\odot$}
\def\lsol{L$_\odot$}
\def\arcsecs{$^\prime \! . \! ^\prime$}

\def\ergscm{erg s$^{-1}$ cm$^{-2}$}

\begin{document}
\null
\vskip -0.7truein
\centerline{Astrophysical Journal -- in press}

\title{Revealing the Photodissociation Region: Hubble Space
Telescope/NICMOS Imaging of NGC 7027}
\author{William B. Latter}
\affil{SIRTF Science Center, IPAC, California Institute of
Technology\footnote{Mailing address:  
SIRTF Science Center, 
California Institute of Technology,
MS 314--6, Pasadena, 
CA 91125};
latter@ipac.caltech.edu
}

\author{Aditya Dayal}
\affil{
Infrared Processing and Analysis Center and Jet Propulsion
Laboratory\footnote{Mailing address:  
Infrared Processing and Analysis Center, 
California Institute of Technology,
MS 100--22, Pasadena, 
CA 91125};
adayal@ipac.caltech.edu
}
\author{John H. Bieging, Casey Meakin}
\affil{
Steward Observatory,
University of Arizona,
Tucson, AZ 85721;
jbieging@as.arizona.edu; cmeakin@as.arizona.edu
}

\author{Joseph L. Hora}
\affil{
Harvard--Smithsonian Center for Astrophysics,
60 Garden Street,
Cambridge, MA 02138-1516;
jhora@cfa.harvard.edu
}

\author{Douglas M. Kelly}
\affil{
Steward Observatory,  
University of Arizona,
Tucson, AZ 85721,
and the University of Wyoming;
dkelly@as.arizona.edu
}
\author{A.G.G.M. Tielens}
\affil{
Kapteyn Astronomical Institute,
PO Box 800,
NL-9700 AV Groningen,
The Netherlands;
tielens@astro.rug.nl
}

\begin{abstract}

We report results from a Hubble Space Telescope (HST) and
Near-Infrared Camera and Multiobject Spectrometer (NICMOS) program to
study the distribution of hot neutral (molecular hydrogen) and ionized
circumstellar material in the young planetary nebulae
\n7027. HST/NICMOS provided very high spatial resolution imaging in
line and continuum emission, and the stability and large dynamic range
needed for investigating detailed structures in the circumstellar
material. We present dramatic new images of \n7027 that have led to a
new understanding of the structure in this important planetary
nebula. The central star is clearly revealed, providing near-infrared
fluxes that are used to directly determine the stellar temperature
very accurately (T$_{\star}$ = 198,000 K). It is found that the
photodissociation layer as revealed by near--infrared molecular
hydrogen emission is very thin ($\Delta$R $\sim$ 6$\times$10$^{15}$
cm), and is biconical in shape. The interface region is structured and
filamentary, suggesting the existence of hydrodynamic
instabilities. We discuss evidence for the presence of one or more
highly collimated, off-axis jets that might be present in \n7027. The
NICMOS data are combined with earlier Hubble Space Telescope data to
provide a complete picture of NGC 7027 using the highest spatial
resolution data to date. The evolutionary future of \n7027 is
discussed.

\end{abstract}

\keywords{planetary nebulae: general -- planetary nebulae: individual
(NGC 7027) -- ISM: molecules -- ISM: structure -- molecular
processes}

\section{Introduction}

Planetary nebulae have long been thought of as ionized remnants of
circumstellar material ejected by stars on the asymptotic giant branch
(AGB). That somewhat limited view has changed dramatically, as the
neutral component of the remnant circumstellar envelope has been shown
to remain observable very late into the lifetime of many planetary
nebulae (PNe; \eg \cite{din95}; \cite{hug96}; \cite{hor99}; and
references therein). It is now known that many PNe are characterized
by emission from material that is contained in rapidly evolving
photodissocation regions (PDRs; \cite{th85}; \cite{sd89};
\cite{lat93a}; \cite{nat98}). As the central star and nebula evolve,
UV emission from the star increases rapidly. A photodissociation front
moves through the gas and slowly turns the mostly molecular nebula to
predominantly atomic. At this point, the nebula radiates mainly in the
infrared. When the central star becomes hotter than $T_* \approx
30,000$ K, the circumstellar gas quickly becomes ionized and the
planetary nebula shines brightly in visible light. The chemical
properties of gas in PNe and proto-PNe (PPN) are like those of
interstellar PDRs (\cite{lat92}; \cite{lat93a}; \cite{tie93};
\cite{hor94}), which are well modeled (see \cite{ht97}).  The time
over which molecular material can be present in PNe is typically
predicted to be a rather brief period in the lifetime of PNe (\eg
\cite{tie93}; \cite{nat98}). But, such timescales predicted from
current chemical models are limited by observational data and lack of
a clear understanding of the morphology and density structure of these
objects.

\n7027 is perhaps the best studied planetary nebula, in part because
of its proximity (about 900 pc; e.g. \cite{mas89}) and high surface
brightness at all wavelengths. It has a very rich atomic and molecular
spectrum, making it ripe for study at all wavelengths, and it is rich
in physical and chemical information (see, \eg \cite{gra93a};
\cite{cox97}, and references therein). Because of its compact size as
viewed from Earth ($\approx 15$\arcsec), the morphology, particularly
that for the \h2\ emission, of this object has been suggested, but has
not been clearly revealed (see, \eg \cite{gra93a}; \cite{gra93b};
\cite{kas94}; \cite{lat95}). The elliptical ionized core lies at the
center of an extended molecular envelope (\cite{jhb91}). At the
interface between the cold molecular envelope and the hot ionized
region is an apparent ``quadrupolar'' region of strong near--infrared
(IR) rotational-vibrational molecular hydrogen emission
(\cite{gra93a}) that has been shown to be excited by absorption of UV
photons in the photodissociation region (\cite{hor99}).

A detailed understanding of the morphology of planetary nebulae is
important. It has been demonstrated that there is a strong correlation
between the presence of molecular emission from PNe and the observed
morphology of the PNe, such that objects that contain large amounts of
molecular material are bipolar or butterfly nebulae (\cite{hor99};
\cite{kas96}; \cite{hug96}; \cite{hug89}; \cite{zuc88}). Progenitor
mass also correlates with morphological type, such that higher mass
stars appear to produce bipolar or butterfly nebulae (see
\cite{cor94}). This correlation suggests that the higher mass AGB
progenitors, which probably have the highest mass loss rates, produce
dense, long lived molecular envelopes (\cite{hor99}).

The evolution of PDR and PNe, and the correlation between
morphology with molecular content are not fully understood.
Because NGC 7027 is in a rapid and key moment in evolution (the
transition from neutral, predominantly molecular envelope to an
ionized one), it is worthy of serious study into its chemical and
physical properties and its detailed morphology.  The Hubble Space
Telescope and NICMOS provided the high spatial resolution, dynamic
range, and stable background needed to examine the near-IR nature of
NGC 7027. In this paper, we present narrowband images in the \s1
($\lambda = 2.121$ \micron) line of molecular hydrogen and other
filters that trace the ionized core and the nearby neutral
region. These images show with unprecedented clarity the true
structure of NGC 7027. In Sections 2 and 3, we discuss the data, the
apparent morphology, and the presence of a jet (or jets) in the
object. In \S 4 we present a 3-dimensional model for the overall
structure, including the photodissociation region. Section 5 presents
a discussion of the excitation of the nebula and the central star
properties. We then consider our results in light of previous work,
and discuss the evolution of the PDR in NGC 7027.

\section{Observations and Data Reduction}

We have imaged \n7027 with the Hubble Space Telescope (HST) and the
Near-Infrared Camera Multiobject Spectrometer (NICMOS) at 7
near-infrared wavelengths between 1.1 \micron\ and 2.15 \micron\
(Figs.\ 1 and 2).  The 1.10 \micron\ broadband (F110W) observations
were made with Camera 1 (pixel scale = 0.043\arcsec\ pixel$^{-1}$,
field of view or FOV = 11\arcsec$\times$11\arcsec). All of the other
observations were made with Camera 2 (pixel scale = 0.075\arcsec\
pixel$^{-1}$, FOV = 19.2\arcsec$\times$19.2\arcsec). The observations
were made in one of the multiaccum detector readout modes (usually
multiaccum 256; see the NICMOS Instrument Handbook) to obtain a very
high dynamic range. A variety of dither patterns were utilized to
sample the array uniformly; in the case of the F110W observations
dithering/mosaicing was required to obtain a complete image of the
source. Chop (or blank sky) frames for background subtraction were
also obtained for all filters except F110W and F160W (1.60 \micron\
broadband). From ground-based observations it was known that the full
extent of the \h2\ emission in \n7027 is slightly larger than the
NICMOS Camera 2 field of view.  A specific orientation of the
Observatory was used to put the largest known extent of the emission
along a diagonal of the array. Dither patterns were used that would
ensure complete coverage of the emission. The method worked very well,
and no emission was missed, although it does appear that very low
level scattered light does extend off the array in some of the bands
observed. A summary of the observations is presented in Table
\ref{tblsumobs}, and the data are presented in Figs. 1 and 2.

The data were originally reduced and calibrated by the standard Space
Telescope Science Institute (STScI) NICMOS pipeline, but we
subsequently re-reduced all of the data using better characterized
calibration files (flat-fields, dark frames, and bad pixel masks) as
well as a revised version of the STSDAS calibration task
calnicA. After reduction of individual frames, the multiple images at
each wavelength were mosaiced together using the STSDAS calnicB task
in IRAF\footnote{The Image Reduction and Analysis Facility is written
and supported by the National Optical Astronomy Observatories (NOAO)
in Tucson, AZ}. The reduced images were flux calibrated using NICMOS
photometric calibration tables provided by M. Rieke (private
communication). The HST Wide Field/Planetary Camera (WFPC2) data were
acquired from the STScI HST data archive.

\paragraph{Isolating the H$_2$ emission --}
An examination of the near-IR spectrum of NGC 7027 (\cite{hor99}), and
the F212N and F215N filter transmission curves shows that both filters
suffer from line contamination.  In addition to the \h2 \s1 line at
2.121 \micron, the F212N filter also transmits the 2.113 \micron\
\ion{He}{1} line at $\sim$ 83\% of peak transmission.  The F215N
filter, used for sampling continuum emission adjacent to the \h2\
line, includes within the filter bandpass the \Brg\ line at 2.167
\micron\ at $\approx$ 2 -- 4\% transmission. We made use of all
available data, theoretical modeling, and empirical fitting to
provide the cleanest possible subtraction of the continuum and
contaminating lines in the \h2\ image. A simple subtraction of the
flux calibrated continuum image from the line image not only isolates
the \h2 emission but also shows a dark (negative) inner ring-shaped
region that resembles the ionized nebula; this inner region is an
artifact of over--subtraction of the \ion{He}{1} line contribution
from the \h2\ filter data. Typically, a bright ring of \ion{He}{1}
emission is seen in ground--based narrowband \h2\ images of NGC 7027
(see, \eg \cite{gra93a}; \cite{lat95}; \cite{kas94}; \cite{kas96}),
because the continuum filters most often used do not include \Brg\
emission in the bandpass.

Since uncertainties associated with filter transmission are largest
along the edges of the bandpass (where transmission changes rapidly
with wavelength), we first made an estimate of the \Brg\
transmission. Using the F190N image as ``true'' continuum for the
F215N image, we compared the observed \Brg\ flux with that obtained by
\cite{hor99} using ground--based spectroscopy. We measured our fluxes
at both of the spatial locations at which their slits were placed
(labelled N and NW in their paper) and found that we detect
approximately 7 -- 8\% of their \Brg\ fluxes at each point. Since the
\Brg\ to \ion{He}{1} ratio is $\sim$ 15 (\cite{hor99}) and assuming
that the \ion{He}{1} line is transmitted at 83\% of peak, this implies
that we need to scale the F215N image by $\sim$ 0.8 to cancel out the
\ion{He}{1} emission in the (F212N -- F215N) difference image.  This
process of eliminating line contamination assumes that the the
\ion{He}{1} to \Brg\ ratio is fairly constant over the
nebula. Photoionization modeling using CLOUDY (\cite{fer97}) shows
that S$_{\rm {He I}}$/S$_{\rm{Br\gamma}}$ ratio does not change by
more than 20\% over the region 1.5$\times$10$^{17}$ cm to the outer
edge, 3$\times$10$^{17}$ cm ($\approx$ 10\arcsec -- 20\arcsec).  Thus,
we can scale the \Brg\ emission (\ie the F215N image) to cancel out
the \ion{He}{1} emission in F212N -- F215N difference image to a very
high degree.  Given the uncertainties involved in this procedure, we
used scale factors in the range 0.7 -- 1 to determine the best \h2\
difference image empirically.  We found that a value of 0.9, when
applied to the F215N image, provided the cleanest subtraction of the
continuum and \ion{He}{1} from the F212N image. This scaling factor is
consistent with $\sim$ 6\% \Brg\ transmission in the F215N
filter. This procedure has resulted in the nearly pure molecular
hydrogen emission image of NGC 7027 shown in Figure 3.
 
The continuum subtracted \h2 image has an integrated line brightness
of 3.8$\times$10$^{-12}$ erg s$^{-1}$ cm$^{-2}$ and an average surface
brightness of 8$\times$10$^{-4}$ erg s$^{-1}$ cm$^{-2}$ ster$^{-1}$.
Using a similar process of aligning images before subtraction we find
an integrated \Pa\ line flux (the F187N line filter and F190N
continuum filter; see Fig.\ 1) of $\approx 5.5\times$10$^{-10}$
\ergscm (Fig.\ 1).

\section{General Results}

\subsection{Observed Nebular Properties} 

Most of the images (narrow- and broad-band) are remarkably similar
(see Figs.\ 1 and 2).  The emission is dominated by a bright, patchy
elliptical ring (major axis along PA =
150\arcdeg) which is the ionized nebula; the overall shape and
extent of this nebula is similar to that previously observed in
visible light, near-IR, and radio continuum images (\eg \cite{roe91};
\cite{woo92}; \cite{gra93a}). Compared to the visible light images of
HST/WFPC2 that show a ``filled-in'' ellipsoid, the near-IR morphology
(where optical depths are lower) clearly shows an ellipsoidal
shell. The high spatial resolution of these new HST/NICMOS images
clearly resolves the ellipsoidal shell into apparent clumps (or
``clump-like'' variations in brightness), regions of varying surface
brightness, and dark dust lanes. The brightest region appears to be a
clump located about 4\arcsec\ NW of the center of the nebula (marked
by the location of the visible central star). The central star stands
out clearly in the continuum images. The similarity between the
\Paa\ and continuum images suggest that the broad-band filters trace
free-free or free-bound emission from hot gas, rather than
dust-scattered starlight, which is apparent as extended emission in
the WFPC2 data. The F212N filter clearly shows a second, spatially
distinct component that is not seen in the other filters: wisps of
ro-vibrationally excited \h2\ emission that appears to surround
the ionized nebula. Furthermore, there is also clear evidence of a
disruption of the gas along a NW-SE (PA $\sim$ 130\arcdeg\ measured E
of N) axis; this might be caused by wind interactions with an off-axis
bipolar jet (see \S \ref{jet}).

The continuum--subtracted \h2 image (Fig. \ref{h2only}) clearly shows
two bright intersecting rings, which we interpret as brightened rims
of an inclined biconical structure (\S \ref{h2model}). This structure
encloses the roughly elliptical ionized region. The structures seen in
molecular hydrogen emission and in tracers of the ionized core appear
different and spatially distinct.

\subsection{A Possible Collimated, Off-axis Jet}\label{jet}

The \h2\ emission (Figs.\ 1 -- 3) shows bubble-like extensions at the
outer edge of the bulk emission region. This apparent disturbance
falls along an axis $\approx$ 20\arcdeg -- 40\arcdeg\ (PA =
130\arcdeg\ NW to SE) to the west of the polar axis, extending through
the central star to the opposite side of the nebula (PA = 130\arcdeg\
NW to SE). There are hints of these structures in ground-based
near--IR data (\cite{gra93a}; \cite{kas94}; \cite{lat95}) as well as
in radio continuum images of the ionized region
(\cite{roe91}). Because of the clarity of these new data, we can now
suggest a cause for this structure. The wisps of \h2\ emission
revealed clearly in these HST data indicate that this region has been
disturbed by something other than dust--photon interactions or a
uniform, radially directed, outflowing wind. We suggest that there is
present in \n7027\ a highly collimated, ``off-axis'' jet. The
disturbance is not only seen in the molecular region, but is clearly
evident in the ionized core along precisely the same position
angle. The structure extends into the region of scattered emission
seen in the WFPC2 images as well (see also \cite{roe91}). The
sharpness of the extended WFPC2 emission argues against something like
turbulent instabilities. There is no kinematical evidence for the jet,
nor is there any extended collimated emission in the HST images; this
leads us to believe that the jet may have disrupted the nebula in the
past and has now died down.  At a much lower level, there appears to
be similar evidence for another jet, or the same jet precessing, at a
polar angle of $\approx$ 10\arcdeg\ to the east of the north
pole. Along an axis in this direction there is an additional region of
``disturbed'' \h2\ emission most clearly seen on the southern side of
the nebula. The ionized core has apparent breaks or disruptions along
this axis on the northern side, and extensions can be seen in the
WFPC2 data as well (Figs. 2 and 3). The locations of these features
are indicated by lines 1 and 2 in Fig.\ 3. 

We speculate that the disturbed morphology of NGC 7027 might be a
result of one or more BRETs (bipolar rotating episodic jets; \eg
\cite{lop98}).  
In this regard we note that the two
axes that we have identified simply show the most pronounced structure;
however much of the other billowy structure and wisps (especially 
prominent in the H$_2$ data) is likely caused by the same, or similar
processes. 
Evidence for precessing, collimated jets that are on,
or away from the primary PN axis are also seen in a number of other
PNe. The most striking example might be NGC 6543 (the ``Cat's Eye
Nebula;'' \cite{lam97}). \n7027 appears to be in a very early stage of
forming a jet, or perhaps the jet is a transient phenomenon. The fact
that we do not see the jet itself suggests that either we have not
found the right diagnostic line or wavelength to probe the jet
kinematically, or the jet has now shut off.

\section{A 3-dimensional Spatial and Kinematical
Model}\label{3dmodel} 

We have modeled the spatial and kinematical structure of \n7027 using
a geometrical modeling code described by Dayal \etal (1999). The code
uses cartesian coordinates (for adaptability to different geometries),
and the nebula is divided into 10$^7$ 0\arcsecs 075-sized cells. In
these models the nebula is assumed to have azimuthal symmetry with the
star at the center.  Once the shape and dimensions of a source have
been set, the density, temperature, or velocity fields can be
specified independently as functions of the radial distance from the
center and/or latitude. 
The expansion of the nebula is assumed to be purely radial, directed
away from the central star.
The tilt angle of the nebula (w.r.t. the plane
of the sky) can be varied independently for each model.  Assuming that
the emission is optically thin, the code numerically integrates the
emission per cell along each line of sight, to produce surface
brightness and position--velocity (P--V) maps. 
The surface--brightness maps are convolved with a gaussian beam (FWHM =
0\arcsecs 25) to match the instrumental point spread function. 
The source function,
which determines the emission in each cell, can be specified as a
function of the local temperature and/or density depending on the
emission mechanism that is being modeled (recombination line emission,
dust thermal emission, dust scattering etc.).

The physical parameters for the nebular model are well constrained by
the high spatial resolution afforded by HST and NICMOS.  Ground--based
kinematical data that have been obtained with CSHELL at the IRTF
provide radial velocity information, which imposes even stronger
constraints on the nebula's 3-D structure; these data also provide
valuable information on the dynamics and evolutionary timescales of
the nebula.  A full presentation of these data will be made elsewhere
(\cite{kel99})  Note that changing the model geometry
changes the shape of the image and the P-V diagram; changing the
density law changes the relative location and contrast of bright and
dark regions in the image and the P-V diagram, but not the shape/size
of the image; finally changing the velocity law does not affect the
image but changes the shape of the P-V diagram. Therefore, model
images together and P-V diagrams, impose strong constraints, allowing
us to develop fairly robust models for the spatial and kinematical
structure of the source.

We focus here on reconstructing the 3-dimensional structure of the
excited H$_2$ shell, which represents the PDR. However we also
construct a model of the \Pa\ emission and compare the density and
velocity structure of ionized gas that we derive, with \Pa~and radio
recombination lines respectively, and estimate a total mass for the
ionized gas. Finally, we also construct a spherical, isotropically
scattering dust model and compare those results with a WFPC2 continuum
image. Note that each of the 3 components is modeled separately, using
independent constraints provided by different images. We adopt source
functions and emission coefficients appropriate for the nebular
temperature and density, for each of the three components.

\subsection{Excited H$_2$ -- The Photodissociation Region}\label{h2model}

\subsubsection{Geometry}

The \h2\ emission is modeled as a ``capped'' bi-conical
(hourglass-shaped) structure which shares the same axis, but lies
exterior to the ionized nebula (Fig.\ \ref{modelschem}).  The outer
dimensions of the bi-cone are well constrained by the continuum
subtracted \h2\ image (Fig. \ref{h2only}): The intersecting outer
rings represent the maximum diameter of the cone; the major axis to
minor axis ratio constrains the tilt of the cone to be $i \approx$
45\arcdeg$\pm$5\arcdeg\ (w.r.t.\ the plane of the sky). Based on an
established tilt and a maximum cone diameter, the observed separation
between the large rings determines the height of the cone. The radius
of the waist of the hourglass is harder to constrain from the data due
to projection effects. We simply assume that this dimension matches
the minor axis of the ionized gas ellipsoid (\S 4.2), R$_w$ =
2\arcsecs 8, though in reality, the neutral gas probably lies somewhat
outside the ionized gas.  The waist diameter along with the maximum
diameter of the cone allows us to estimate an opening angle for the
cone. There is some evidence in the images that the H$_2$ emission is
not confined to a conical shell that is open at the poles, but extends
all around the ionized gas ellipsoid. Wisps of \h2 emission are seen
superposed on the northern and south-eastern parts of the ionized
nebula in Fig. 2. Also, our model images show that an ``open'' cone
cannot produce the bright rims, along the top and bottom of the
structure, as are seen in the images.  When a ``cap'' of similar
thickness as the walls of the bicone is added, limb brightening along
lines of sight through the caps alleviates this problem (see Figs. 5(a),(b)).
By comparing intensity cuts across the model with similar cuts across
the image we more accurately determine the dimensions of the H$_2$
shell.  The walls of the bicone are constrained by the thickness of
the observed H$_2$ rims, $\Delta$R $\sim$ 0\arcsecs 5$\pm$0\arcsecs 1
(6$\pm1\times10^{15}$ cm at the adopted distance of 0.88 kpc). For
example, increasing the width of the shell reduces the gap between the
rims of the H$_2$ bicone and the inner, ionized gas ellipsoid;
decreasing the width of the shell produces narrower looking rims than
are seen in the NICMOS-only color composite image (Fig. 2(a)). Because
the nebula is inclined to the line of sight, it is difficult to
ascertain variations in the thickness of the H$_2$ shell as a function
of latitude (particularly at low latitudes). We assume therefore that
the shell has a constant width. 

\paragraph{Limitations --} The cone/cap model here is an
approximation to the nebula, which probably has a smoother, bi--lobed
shape. However, building up the model in separate components allows us
to see the effects of the cap (or lack thereof) on the P-V diagram and
images. Though we do not believe that a smoother hourglass figure model
would yield significantly different results, it may match the brightness
in the interior (equatorial regions) of the H$_2$ images somewhat better  
than our current models, which appear to overestimate the H$_2$ 
brightness in this region.

\subsubsection{Kinematics}

Long-slit CSHELL spectra of the \h2\ \s1 line are invaluable in
constraining the geometry of the source more tightly. The
P-V diagram obtained by placing the slit along the
major axis of the source is shown in Fig.\ \ref{posveldig}. The
emission region has an elongated elliptical shape suggesting that
there are distinct red and blue--shifted components of emission both
in the top and bottom halves of the nebula.  The shape of the P-V
diagram confirms that the nebula is oriented such that the N-E part of
the shell is tilted {\it towards} the observer and the S-W part is
tilted {\it away} from the observer.  This orientation is consistent
with that of the ionized gas ellipsoid observed by \cite{bai97}, who
used an echelle spectrograph with a velocity resolution of 6 km
s$^{-1}$.  The four bright ``spots'' of emission, located
symmetrically about the center of the nebula, indicate locally bright
regions along the slit separated in velocity. The two bright regions
closer to the center mark the position of the limb--brightened inner
rims of the bicone on the slit, while the bright regions at the top
and bottom correspond to the limb-brightened ``caps'' of the bicone,
which lie almost in the plane of the sky.  It is the position-velocity
diagram that uniquely constrains the distribution of H$_2$ along the
poles. Figs. 5 (c) and (d) show model P-V diagrams obtained with and
without a cap. Clearly a conical structure {\it with} the spherical
cap gives a more complete fit to the observed P-V diagram
than one without the cap.  A constant
expansion velocity of 20 km s$^{-1}$ is able to reproduce the observed
spectra quite well though we cannot rule out deviations from this law
on smaller spatial scales. Figs. \ref{posveldig} (b), (c) show how the
shape of the model P-V diagram changes with the velocity law. In
general, a larger velocity law exponent results in a larger ``kink" in
the position velocity diagram, corresponding to more abrupt change in
the projected velocity from the rim of the cone (large R) to the
interior of the cone.  A comparison with the data allows us to rule
out a velocity exponent, $\alpha \geq 1$. Thus, for our adopted model
geometry we can rule out a constant dynamical timescale for all points
in the \h2~shell, as one might expect for an aspherical shell
expanding in a self-similar way.  The velocity law implies that at a
latitude of 25\arcdeg~the walls have a dynamical timescale of 950 yr
and further out, at 36\arcdeg\ (where the bicone ends and the
spherical cap begins) the timescale is $\sim$ 1,300 yr. We note that
since this is an expanding photodissociation front, the dynamical
timescale is not the most appropriate one for characterizing the
evolution (see \S 6.2).

\subsubsection{Density}

Though the excitation of H$_2$ is primarily via absorption of UV
photons (i.e. non-thermal, see \S 5.2) we assume here that the \h2 is
thermally excited and that the level populations are described by a
vibrational excitation temperature in the range $T_{vib} \approx 6,000
- 9,000$ K (see \cite{hor99}).  Later in this paper we show that this
assumption gives reasonable results when compared to more detailed
models of dense PDRs.  If the H$_2$ emission is optically thin, then
the model intensity at each pixel is simply proportional to the
density of H$_2$ molecules integrated along a given line of sight:
\begin{equation}
\epsilon_{H_2} = \int j_{H_2} n_{H_2} \,dl
\end{equation}
where the emission coefficient (in erg s$^{-1}$ cm$^{-2}$ per molecule) is
\begin{equation}
\epsilon_{H_2} = \frac{A_{u,l} g_N (2J+1) h\nu}{4\pi Q(T) e^{\frac{E_u}{T}}  } 
\end{equation}
and $dl$ is the spatial increment along the line of sight.  The
spontaneous radiative decay coefficient for the 1--0 S(1) line,
$A_{u,l}$ = 3.47$\times$10$^{-7}$ s$^{-1}$ (\cite{wol98}), the nuclear
spin statistical weight g$_N$ = 3 (ortho-H$_2$) and J=3. The
energy of the upper level of the transition, E$_u$ = 6947 K. Using the
polynomial coefficients for the partition function, $Q(T)$, given in
\cite{sau84} we calculate the change in the product, $\delta$ $\equiv$
Q(T) e$^{\frac{E_u}{T}}$ with temperature in the range 2000 K $\leq$ T
$\leq$ 11,000 K. We find that $\delta$ changes by about a factor of
1.5 over the range 6,000 $\leq$ T $\leq$ 9,000 K. We simply assume
$\delta$ = 220, and use this in the the above equation.

Using the equation above to fit the observed integrated intensity of
H$_2$ (3.8$\times$10$^{-12}$ erg s$^{-1}$ cm$^{-2}$; \S 3.1.1) and the
H$_2$ radial brightness profile (Fig. 7(a)) allows us to estimate an
{\it excited} H$_2$ density of $n \sim$ 10--13 cm$^{-3}$ (depending
upon the poorly known attenuation value) for a filling
factor of unity. Summing over the entire nebula then yields a total
mass of $M_{ex} \approx 10^{-5}$ \msol\ for the excited H$_2$. A
constant density H$_2$ law provides adequate fits to the images and
the P-V diagrams.

\subsection{Ionized Gas Emission and Scattered Starlight}\label{ionmodel}

\paragraph{\Pa~emission --}

The ionized nebula is well reproduced by a prolate ellipsoidal shell,
as has been shown previously (\cite{ath79}; \cite{roe91}).  Using the
inclination of 45\arcdeg\ estimated from the \h2\ image, we calculate
the lengths of the semi--major and semi--minor axes (a$_y$ and a$_x$
in Fig. 4).  The thickness of the shell (though it has a broken,
patchy appearance) is estimated to be about
0\arcsecs 9 (1.2$\times$10$^{16}$ cm), from the radial cuts taken in
orthogonal directions through the walls (see Table 2).  Ignoring the
contribution of electrons from heavier elements (i.e. assuming n$_e$ =
n$_H$), the model surface brightness in \Pa\ is given by:
\begin{equation}
I_{P\alpha} = \int j_{P\alpha} n_e^2 \,dl
\end{equation}
where j$_{P\alpha}$ (\Pa\ emission coefficient) = 
3.27$\times$10$^{-27}$ erg cm$^3$ s$^{-1}$ ster$^{-1}$ assuming case
B recombination,  T$_e$ = 10$^{4}$ K (Table 4.4; \cite{ost89}) and l is 
the path length through the nebula at any given position 
on the projected surface of the nebula.

The density law for the ionized gas is constrained by comparing the
model image (and radial cuts) with the \Pa\ image, although the clumpy
nature of the ionized shell makes it somewhat difficult to fit a
well--defined power law. Our results indicate that models with a
constant density or slowly decreasing radial power laws
(R$^{-\alpha}$, 0 $\leq \alpha \lesssim$ 0.5) models provides an adequate
fit to the data (Figs. 7(b)). In models with $\alpha$ $\geq$ 1 the
polar densities are considerably lower than those along the equator
(Figs. 8(c),(e)); consequently radial cuts along the poles show a
smaller rise in intensity going from the center outwards, than do
radial cuts along the equator. The cuts through the observed image
clearly do not show such a difference between the orthogonal radial
cuts. 

We are able to match the observed \Pa~image with a constant density
model where the density in the shell is
n$_{H^+} = 6\times$10$^{4}$ cm$^{-3}$. We also impose an upper limit
to the density inside the ellipsoidal shell, n$_{e,interior}$ $\leq$
5$\times 10^{3}$ cm$^{-3}$.
These densities, together with the derived
geometry imply that the total ionized mass is $M_{ion} \sim$ 0.018
\msol. (The mass might be about 50\% higher if the attenuation factor
is, A$_V$ = 2.97 mag -- see Section 5.1). This low density inner cavity 
implies high temperatures
($\gtrsim 1 \times 10^5$ K), possibly shock heated in interacting
winds. This suggests that diffuse X-ray emission from a hot gas might be
detectable from NGC 7027, in addition to point source X-ray emission
from the central star.

Our derived shell density agrees very well with the density derived by
Roelfsema \etal (1991; 6$\pm0.6\times10^4$\ cm$^{-3}$) from radio
continuum observations. We are unable to reproduce the total ionized
mass of 0.05 \msol~that they infer, given their ellipsoid dimensions
and density (at an assumed distance of 1 kpc). Using their parameters
we find an ionized mass, M$_i \sim$ 0.025 \msol.
The PV diagrams presented in Fig. 8 (which assume that velocity is constant)
are consistent with that
presented for ionized gas by \cite{roe91}. A careful examination of 
the intensity contrast in the constant--density model suggests that
this model most closely matches these data, given the S/N of the data.

\paragraph{Dust scattering --}\label{dustmodel}

An examination of the WFPC2 F555W image of NGC 7027 shows that the
ionized nebula is surrounded by an extended region of lower surface
brightness. The radial profiles indicate that the intensity falls off
as I $\propto$ R$^{-3}$, the expected index for singly scattered
starlight from an optically thin region of the dust shell. We model
the scattered starlight as single (isotropic) scattering from a
spherical shell of 0.1 \micron~sized amorphous carbon grains, with a
scattering cross-section, K$_s$ = 8$\times$10$^4$ cm$^2$ gm$^{-1}$ at
0.55 \micron~(\cite{mar87}). The photon source is assumed to be a
200,000 K black-body. Assuming an AGB dust mass loss rate of $10^{-6}$
\msol~yr$^{-1}$ and an expansion velocity of 10 km s$^{-1}$ yields a
model flux density of $\sim$ 4.6$\times$10$^{-4}$ Jy arcsec$^{-2}$ at
a distance of 5\arcsec~ from the center of the nebula. The model radial
brightness profile appears to match the observed surface brightness 
in the F555W image, along the minor axis of the nebula (Fig. 7c). The
dust mass loss rate agrees well with the gas mass loss rate of
1.5$\times$10$^{-4}$ \msol~ derived by \cite{jam91} if the gas to dust
ratio is $\sim$ 150.

We have tried to fit the broad--band near--IR emission with the same 
scattering model as used above for the WFPC2 image. We find that assuming
$\sigma_s$ $\sim$ $\lambda^{-4}$ (Rayleigh scattering, appropriate for
small grains) yields near--IR intensities that are 
considerably lower than those observed in the F160W and F205W filters. 
A better estimate is obtained if the scattering cross--sections goes
as $\lambda^{-1}$, implying that the dust scattering is dominated by
larger (``fluffy") grains. Also, the radial profiles in the near-IR filters
do not show a well--defined radial power law as seen in the WFPC2 image,
suggesting that the optically thin, singly scattered starlight model is 
inadequate in describing the extended near-IR emission.
We suspect that the emission in the near--IR broad--band filters is not
simply scattered starlight but includes scattered line/f-f emission as well
as possibly some dust thermal emission. A quantitative description of 
this ``excess" near--IR emission will require a detailed model of the
gas and dust emission and is not addressed further this paper.

\section{Central Star and Nebular Excitation}

\subsection{Properties of the Central Star}\label{cs}

The central star (CS) of NGC 7027 has been detected only with
difficulty from the ground at visible wavelengths and under average
seeing, because of the low contrast of the star relative to the bright
nebular emission (Jacoby 1988; Heap and Hintzen 1990).  NICMOS offers
high resolution imaging at longer wavelengths where extinction from
interstellar and circumstellar dust is much smaller than at visible
wavelengths, and the contrast of the stellar point source is greatly
enhanced by the high angular resolution of HST.  This observational
advantage allows a significant improvement in photometric data for the
CS and in the stellar properties thereby inferred.
  
Photometry for the CS was obtained for the F110W, F160W, F190N, and
F205W NICMOS filters.  We have also determined CS fluxes for the F555W
and F814W filters of the WFPC2 optical camera, from images in the HST
data archive.  Nebular photometry of the P$\alpha$ and \h2\ \s1\ line
emission was derived from the line and continuum filter sets, F187N
(line) and F190N (adjacent continuum), and F212N (line) and F215N
(continuum) respectively. The photometry of the central star and the
nebular line emission is summarized in Table 3.
 
The largest source of uncertainty in our stellar photometry arises
from the spatially variable attentuation and the complex, clumpy
structure of the nebula.  Near the CS, this structure varies on a
scale as small as the radius of the first minimum in the PSF, so
background estimation and subtraction are difficult.  We found that
the best NICMOS photometry is achieved by subtracting a scaled
P$\alpha$ line emission image from each of the images on which
photometry is being performed.  The scale factor is chosen to make the
nebular background have a zero mean in the vicinity of the CS.  (The
P$\alpha$ image is obtained by subtracting the F190N continuum image
from the F187N P$\alpha$ narrow band image, so that the central star
is removed, leaving only the nebular line emission.)  In this
procedure we assume that the P$\alpha$ line emission is a good tracer
of total nebular line and continuum emission in the other NICMOS
filters, at least in the region near the CS.  We find that this method
successfully removes the spatially variable nebular emission near the
CS.

In the WFPC2 images the nebular surface brightness shows smaller
variations relative to the CS.  In this case we employ more
conventional  photometry and assume that the surface brightness in an
annulus immediately  surrounding the CS is a fair representation of the
background. The photometry was converted to flux density
by applying the photometric calibration factor (PHOTFLAM) provided by
STScI in the image header.

The CS photometry is dereddened using an attenuation value at the
position of the CS of $A_V = 3.17$ mag. The P$\alpha$ line emission
photometry is corrected for an average nebular attenuation of $A_V =
2.97$ mag (Robberto et al.\ 1993).  The correction at each wavelength is
derived from the mean extinction curve given in Table 3.1 of Whittet
(1992).
 
We calculate a hydrogen Zanstra temperature (cf.\ Osterbrock 1989) for
each CS photometric value in Table 1, with the assumption of a
blackbody CS spectrum.  For the nebular line emission, we use the
integrated P$\alpha$ line flux given in Table 3.  The greatest source
of error in this calculation results from uncertainties in the nebular
conditions on which the recombination coefficients depend.  We adopt a
nebular electron temperature of 15000 K and a nebular electron density
of 10$^4$ cm$^{-3}$ (cf. Roelfsma et al. 1991; Masson 1989), and use
the corresponding Case B total recombination coefficient, $\alpha_B$,
and P$\alpha$ emissivity calculated by \cite{sto95}.  The derived
Zanstra temperatures of the CS are compiled in Table 3.

The derived CS temperatures are remarkably consistent (see Table 3).
We find $T_*=198,100\pm 10,500$ K.  Adopting a
distance of 880$\pm$150 pc (Masson 1989), we find from the photometric
data of Table 3 a mean CS luminosity of 7710 $\pm$ 860 L$_{\odot}$,
and a photospheric radius of $5.21\times 10^{9}$ cm.  If we place the
CS on the evolutionary tracks for hydrogen-burning post-AGB stars of
Bl\"{o}cker (1995), the CS lies very close to the theoretical track
for an initial mass of 4 M$_{\odot}$ and a final core mass of 0.696
M$_{\odot}$.  If this model is appropriate, the CS would have left the
AGB only about 700 years ago.  This timescale is consistent with the
estimated age of 600 years for the ionized nebula (Masson 1989) based
on the expansion velocity and size. The stellar evolutionary timescale
is also comparable to the dynamical timescale of the \h2\ lobes
estimated in \S 4.1.2 above. Jaminet \etal (1991) have observed the
neutral molecular envelope which surrounds the ionized gas and
determined a mass of 1.4 (+0.8, --0.4) M$_{\odot}$.  If the CS initial
mass was in fact 4 M$_{\odot}$, and the final core mass is 0.7
M$_{\odot}$, then the star must have shed 3.3 M$_{\odot}$ over the
course of its evolution.  Of this amount, 40 -- 50\% was lost in the
last 10$^4$ years, when the mass loss rate was some $1.5 \times
10^{-4}$ M$_{\odot}$ yr$^{-1}$ as inferred from models for the CO
emission of the envelope (Jaminet et al. 1991).  The balance of the
mass loss might have occurred more slowly over a much longer time
period, but the lack of clear observable signatures of this matter
make it impossible to infer the rate or timescale of this earlier
phase with any precision. The stellar properties found for NGC 7027
are summarized in Table 4.

\subsection{Molecular Hydrogen Excitation and the Morphology of
\n7027}\label{h2e} 

The near-infrared spectrum of molecular hydrogen is the result of slow
electric quadrapole transitions within the ground electronic state
vibration-rotation levels. A possible excitation mechanism
likely in PNe is via the absorption of far-ultraviolet (FUV) photons in the
Lyman and Werner bands of \h2. Upon absorption of a FUV photon, the
molecule is left in an excited electronic state from which $\approx
$10\% of the decays out of that state result in transitions to the
ground electronic state vibrational continuum and molecular
dissociation occurs. Most often the molecule is left in an excited
vibration-rotation level in the ground electronic state. The result is
an identifiable near-IR fluorescence spectrum. This process is the
primary way \h2\ is destroyed, by photons and the peak in
ro-vibrationally excited molecules defines the \h2\ photo-dissociation
region (PDR). In addition, the excitation of the near-IR \h2\ spectrum
can occur in regions with kinetic temperatures of $T_K \gtrsim 1000$ K
and total densities $n_{tot} \gtrsim 10^4$ \cm3. In PNe such
temperatures are typically thought to be associated with shockwaves of
moderate magnitude and over a limited range ($V_s \approx 5 - 50$
\kms; with the higher end of the range for $C$-type shocks only;
\cite{hol77}; \cite{dra82}).

Graham \etal (1993a) argued, primarily from morphology, that the
near-IR \h2\ emission from \n7027\ is caused by FUV excitation in the
PDR. An analysis of near-IR spectra of \n7027\ (\cite{hor99}) confirms
that hypothesis. There is no evidence of shock excitation, though
localized shocked regions cannot be ruled out.

In a PDR the \h2\ emission is observed in a thin transition region
where the \h2\ column density of FUV excited molecules peaks. Interior
to the transition region the molecules have mostly been
dissociated. Because the excitation process is by absorption of line
photons, the depth of excitation is strongly limited by self-shielding
of the molecules to Lyman and Werner band line radiation. The total
extent of the transition region depends on molecular and total
density, gas to dust ratio, and the strength and shape of the
radiation field. 

Comparison with quasi-steady state PDR models (Latter \& Tielens, in
preparation) finds that an observed average \h2\ \s1\ line brightness
of $8.0\times 10^{-4}$ erg s$^{-1}$ cm$^{-2}$ ster$^{-1}$ (\S 3.1) is
consistent with a C-rich PDR exposed to a single star of temperature
200,000 K and a moderate total density of $n_{tot} \approx 5\times
10^4$ cm$^{-3}$. The predicted $v = 2 - 1$ S(1) to $v = 1 - 0$ S(1)
line ratio ranges from 0.12 to 0.06 and is nearly identical to that
found by analysis of the \h2\ spectrum (\cite{hor99}). For a core mass
of 0.696 \msol\ and the observed \h2\ line brightness, time dependent
models of PDR evolution in PN suggest a PDR lifetime of $t \lesssim
1000$ years (\cite{nat98}) consistent with estimates of the time since
the object left the AGB (\S 5.1). The models are also consistent with
a thin transition (\h2\ emitting) region like that found from the 3-D
models discussed in \S 4. The nonuniformity and patchiness of the
transition region must be a result of structure in the circumstellar
material itself. Instabilities of the type required are not understood
in an interacting wind, photon-dominated environment. The ``waves''
present on the illuminated interior of the \h2\ emission region
resemble those seen in the HST/WFPC2 images of MyCn 18 (the ``Etched
Hourglass Nebula;'' \cite{sah99}; \cite{day99}), and although are seen
in very different components of the gas (ionized versus molecular)
most likely are a result of a similar instability.

It is important to consider whether the overall shape of the PDR (as
it extends from the ionized core to the biconical, or bubble shaped
\h2\ emission) as found from our HST/NICMOS images is consistent with
what is expected for an evolving PDR in a distribution of material
likely to be present in a post-AGB circumstellar envelope. First, we
must determine what is a likely distribution of circumstellar
material. It has been known for some time that post-AGB objects and
PNe generally do not have a spherical (radially symmetric)
distribution of gas and dust. The most extreme examples include AFGL
2688 (the ``Egg Nebula''), AFGL 618, M 2-9, and M 1-16 (to list just a
few; see, \eg \cite{sah98}; \cite{lat95}; \cite{lat92}; \cite{hor94}
and references therein). These objects have a predominantly axially
symmetric distribution of circumstellar material such that the density
is greatest at the equator and lowest near the two poles. With
support from general morphological studies, hydrodynamic modeling, and
polarimetric observations (e.g. \cite{bal94}; \cite{tra94};
\cite{bal97}) this type of distribution (to a widely varying degree)
appears to be generally true for most, or all post-AGB
objects. \n7027\ is not an ``extreme'' bipolar nebula (of the type
listed above), but we can assume that the equatorial to polar density
contrast as the star evolved off the AGB has a functional form that
goes roughly like $\sin(\Theta)$ with additional factors of order
unity. Functions like this have been used to model the scattered light
and molecular distribution from a number of bipolar PPNe and PNe
(\cite{yus84}; \cite{lat92}; \cite{lat93}).

In an environment with a single, central FUV source and with a
constant gas to dust ratio, the dominant parameters that will
determine the radial structure of the PDR are the density structure,
$\rho (r)$ and the distance to the central source. For guidance some
new computations were made using a code developed for a related
purpose, and this topic will be explored quantitatively elsewhere
(Latter \& Tielens, in preparation). To first order, a PDR in a lower
density region will have the ionized/neutral interface farther from
the central source when compared to a higher density
region. Similarly, the region of strongest \h2\ FUV excitation will be
farther out and have a somewhat larger linear extent than in a higher
density region. This H/\h2\ interface region will approximately follow
iso-column density contours at $\sim 4\times 10^{21}$ cm$^{-2}$ or
$A_V \approx 2$ magnitudes. Details will depend strongly on the exact
distribution of material. But, it is evident that an axially symmetric
distribution of circumstellar material of the type discussed above
will produce a PDR morphology like that observed in \n7027. More
specifically, it is expected to be an ellipsoidal or bipolar
H$^+$/H$^\circ$ interface, surrounded by a bipolar bubble structure
defining the H/\h2\ interface and revealed in FUV excited \h2\
emission -- like that modeled here. Somewhat farther out, the C/CO
interface region is, as shown by the data of \cite{gra93a}, consistent
with that found for the inner regions. The morphological evolution of
the H$^+$/H$^\circ$ interface region and the ionized core is not
predicted by strict PDR evolution alone. The internal density
structure appears to have equilibrated in the shell and cavity (\S
4.2) because the sound crossing time is much shorter than the
expansion timescale. The outward motion of this interface will be
faster in the lower density polar regions, magnifying the asymmetric
shape. If the star is on the cooling track, then the ionization front
might better be described as a ``recombination'' front as the number
of ionizing photons goes down (Schoenberner 1999, personal
communication). 

Nested hourglass shapes and ellipsoids are predicted by numerical
hydrodynamical simulations (such as \cite{mel95}; \cite{icke88}), but
for interaction speeds much higher than so far seen in NGC 7027. In
addition, the molecular hydrogen emission from NGC 7027 is from a
strong PDR, not a shocked wind interface. There is similarity in the
overall morphology of the PDR, including the ionized core in NGC 7027
to emission morphologies predicted in strong interacting winds
starting in an axisymmetric density distribution. Because of this
apparent similarity, a full description of the observed structure and
its evolution requires coupled time-dependent PDR and hydrodynamic
modeling. When combined with detailed chemical modeling of the neutral
regions, the data presented here can be used to determine the exact
density distribution of the circumstellar envelope.

\section{Discussion}

\subsection{Previous Kinematic Results}

In a detailed multiwavelength study, Graham \etal (1993a) found that
uniform, radial expansion of a prolate spheroid, tilted to the line of
sight is consistent with the millimeter-wave (CO) position-velocity
maps. 
However the CO P-V diagram differs from the H$_2$ P-V diagram presented
here. This is not unexpected since the observed emission from the two molecules 
is produced under very different excitation conditions. Therefore the two
molecules trace different spatial/kinematical structures. 
Observations have been made with the ``BEAR'' instrument (a
256$\times$256 pixel HgCdTe camera coupled to a Fourier transform
spectrometer, with a spectral resolution \about\ 52.2 \kms) on the
Canada-France-Hawaii Telescope (\cite{cox97}). Cox \etal found that
the velocity field appeared to them as an equatorial ``torus''
suggests that north pole is tilted away from observer. This result is
inconsistent with our interpretation of the HST images, our 3-D
modeling of those images, and the higher spectral resolution CSHELL
data. The high spatial resolution and sensitivity of the HST images
shows that the front part of the top rim lies in front of the the
ionized nebula (not behind), suggesting that the top (northern side)
is tilted $toward$ the observer. We cannot resolve this discrepancy
with the data available. We suggest that the spectral resolution
available to Cox \etal (1997) was not sufficent to fully deconvolve
the subtle, but complex velocity structure that must be present
(\cite{kel99}). In addition, the model schematic given by Jaminet
\etal (1991) shows the same orientation as Cox \etal (north pole away)
for the fast molecular wind traced by CO $J = 3 \to 2$ line emission,
which presumably has a similar flow direction as the ionized gas. It
is entirely possible that the flow along the back of the cone (or
bubble) has been confused with a flow along the polar axis, which in
our model would not be the dominant emitting region. Bains \etal (1997)
found from their high spectral resolution radio and optical
observations, an orientation for NGC 7027 that is consistent with
ours: i.e. that the NE lobe is blue-shifted, SW lobe is red-shifted.

\subsection{Evolution of the PDR in \n7027}

A full modeling of PDR evolution is beyond the scope of this paper. We
can, however, make an assessment of such PDR evolution based on
available models. The morphological evolution of the PDR will depend
most strongly on the distribution of material in the circumstellar
envelope. For \n7027\ the post-AGB circumstellar envelope must have
been axially symmetric with a gradual decrease in density from the
equatorial region to the poles. A dense equatorial disk cannot be
ruled out, but there is no evidence in these data, or other molecular
data (\cite{jhb91}; \cite{jam91}; \cite{gra93a}), that requires or
even suggests one. For that reason, we consider such a structure as
unlikely to be present. The density distribution appears typical for
other post-AGB objects with an equatorial to polar density contrast of
$\sim 2 - 10$, such that it is not as great as the extreme bipolar
proto-PNe (see \S \ref{h2e}). In much less than the cooling timescale
of the central star, the PDR will be established quickly in all
directions, with the fastest evolution in the polar regions. The
result will most likely be for \n7027\ to become a butterfly-type
nebula similar in gross properties to (e.g.)  NGC 2346. Wind
interactions for this type of morphological evolution are not
required.

The timescale for PDR evolution is highly uncertain (other timescales
are discussed in \S \ref{cs}), but reasonable estimates can, and have
been made (see, \eg \cite{tie93}). It is straightforward to show that
for a constant mass loss rate and parameters typical of high mass loss
rate carbon stars, the timescale for a PDR to completely move through
the circumstellar envelope can be estimated by integrating the
equations of molecular formation and destruction with respect to
radius and time, as was done by \cite{tie93}. For a constant wind
velocity $V_w = 20$ \kms, a constant mass loss rate $\dot M = 10^{-4}$
\msol\ yr$^{-1}$, and a stellar luminosity $L_* = 6000$ \lsol, it can
be shown that the dissociation front travels a distance $r_i \approx
3\times 10^{16}$ cm in $t_i \approx 65\ {\rm years}$ with a PDR speed
of $V_i \approx 263\ {\rm km\ s}^{-1}$ (see \cite{tie93} for a more
detailed discussion; see also \cite{nat98}). Tielens suggested
clumping as how the evolution is slowed in real nebulae. Although
widespread clumping and structure is seen in these data, it does not
appear at the size scales required, or marginally so at best ($R
\gtrsim 10^{16}$ cm and $A_V \gtrsim 4$ mag). Even if clumping is
important in localized regions, it is more likely that we now
know several of the assumptions used to derive this evolutionary
timescale are not valid, and if properly accounted for will tend to
increase the timescale. The mass loss rate is not constant. The
``ring'' structure clearly visible in AFGL 2688 (\cite{lat93};
\cite{sah98}) indicates a time varying mass loss rate on a timescale
of hundreds of years. The same type of structures are seen in other
objects observed (\cite{kwok98}), including \n7027 (see, \eg
\cite{bond97}). In addition, the distribution of material is not
spherical. The degree to which the mass loss rate is varying is not
known. Nor is how a non-spherical distribution of attenuating and
shielding material will alter the evolutionary timescales (see,
however, \cite{nat98}). Additional work must be done to characterize
these properties and how they relate to evolution. From the above
expression and an assessment of the impact of uncertain parameters, it
seems likely that the timescale for \n7027\ to evolve away from its
current state will be only a few hundred to a few thousand years (see
also \cite{nat98}). Wind interactions will complicate the picture by
adding hydrodynamic effects and changes to the density
structure. However, evolution to the current epoch appears to be
described fully by the interaction of FUV and ionizing photons with an
AGB wind that has (or had) a somewhat enhanced mass loss rate in the
equatorial plane. While there is evidence for recent jet interactions
with the outflow, the jets have not dominated the morphological
shaping. It is worth monitoring \n7027\ at very high spatial
resolution for morphological changes caused by jets and UV photons
over the next several decades.

\section{Summary}

The HST/NICMOS observations presented in this paper reveal the
detailed structure and morphology of the young planetary nebula NGC
7027 with unprecedented clarity. The molecular photodissociation
region is found to be of an apparently different overall structure
from that of the ionized core. We have constructed 3-dimensional,
axisymmetric models for the morphology of NGC 7027 based on these
NICMOS near-infrared data and HST archive visible light data. The
object can be well described by three distinct components: an
ellipsoidal shell that is the ionized core, a bipolar hourglass
structure outside the ionized core that is the excited molecular
hydrogen region (or the photodissociation region), and a nearly
spherical outer region seen in dust scattered light and is the cool,
neutral molecular envelope. It is argued that such a structure is
consistent with one which would be produced by a photodissociation
region with a central source of far-ultraviolet photons surrounded by
an axisymmetric distribution of gas and dust that has a decreasing
total density with increasing angle toward the polar regions of the
system.

The central star is clearly revealed by these data. We are able to
determine the most accurate value for the temperature of the star $T_*
= 198,100 \pm 10,500$ K with a luminosity of $L_* \approx 7,700$
\lsol. From various indicators, including the central star timescale,
dynamical timescale, and timescale for evolution of the
photodissocation region, we conclude that NGC 7027 left the asymptotic
giant branch 700 to 1000 years ago. For such rapid evolution, it is
likely that objects like NGC 7027 do not go through a proto-planetary
nebula phase that is longer than a few tens of years, and as such
would be nearly undetectable during such a brief phase.

There is strong evidence for one, and possibly two highly collimated
bipolar jets in NGC 7027. The jets themselves are not seen with the
present data, but the disturbance caused by them is clearly visible. It
is possible that this jet, or jets have shut off during the current
epoch of evolution. Such jets have been seen in other planetary
nebulae, but NGC 7027 might be the youngest in terms of
evolution. There are nonuniformities and wave-like structures seen in
these data that must be caused by poorly understood instabilities in
the outflowing wind.

We conclude that while wind interactions will be important to the
future evolution of NGC 7027, it is the evolution of the PDR, as the
central star dissociates, then ionizes, the circumstellar medium, that
will dominate the apparent morphological evolution of this
object. This is in contrast to the idea that interacting winds always
dominate the evolution and shaping of planetary nebulae. Evolution
driven by far-ultraviolet photons might be more common in planetary
nebulae than has been previously thought (see, e.g.,
\cite{hor99}). Such photon-driven evolution will be more important in
objects with higher mass central stars (and therefore high luminosity,
temperature, and UV flux), than in those with low mass stars. But, it
is the high mass objects that also have the higher density molecular
envelopes ejected during a high mass loss rate phase on the asymptotic
giant branch, and as such will show the strongest emission from a
photodissociation region. NGC 7027 is perhaps the most extreme example
known at this time, but it is not unique -- except for the important
and very brief moment in evolution at which we have found it.

\acknowledgments 

This work is dedicated to the memory of Christopher Skinner, our
Contact Scientist at STScI, our colleague, and our friend. We miss his
support, comments, and criticism. He challenged us to search for the
truth, and through that he enriched us all.  

We gratefully thank the members of the NICMOS Instrument Development
Team at the University of Arizona for help, suggestions, comments, and
data; especially Marcia Rieke, Dean Hines, and Rodger Thompson. We
also thank Sun Kwok and Nancy Silbermann for useful conversations, and
Tom Soifer for reading the manuscript. 
Support for this work was provided by NASA through grant number
GO-7365-01 from the Space Telescope Science Institute, which is operated
by AURA, Inc., under NASA contract NAS5-26555. W.B.L. and
J.L.H. acknowledge additional support from NASA grant 399-20-61 from
the Long Term Space Astrophysics Program.

\newpage
%
%
\begin{figure}
\epsscale{0.7}
\plotone{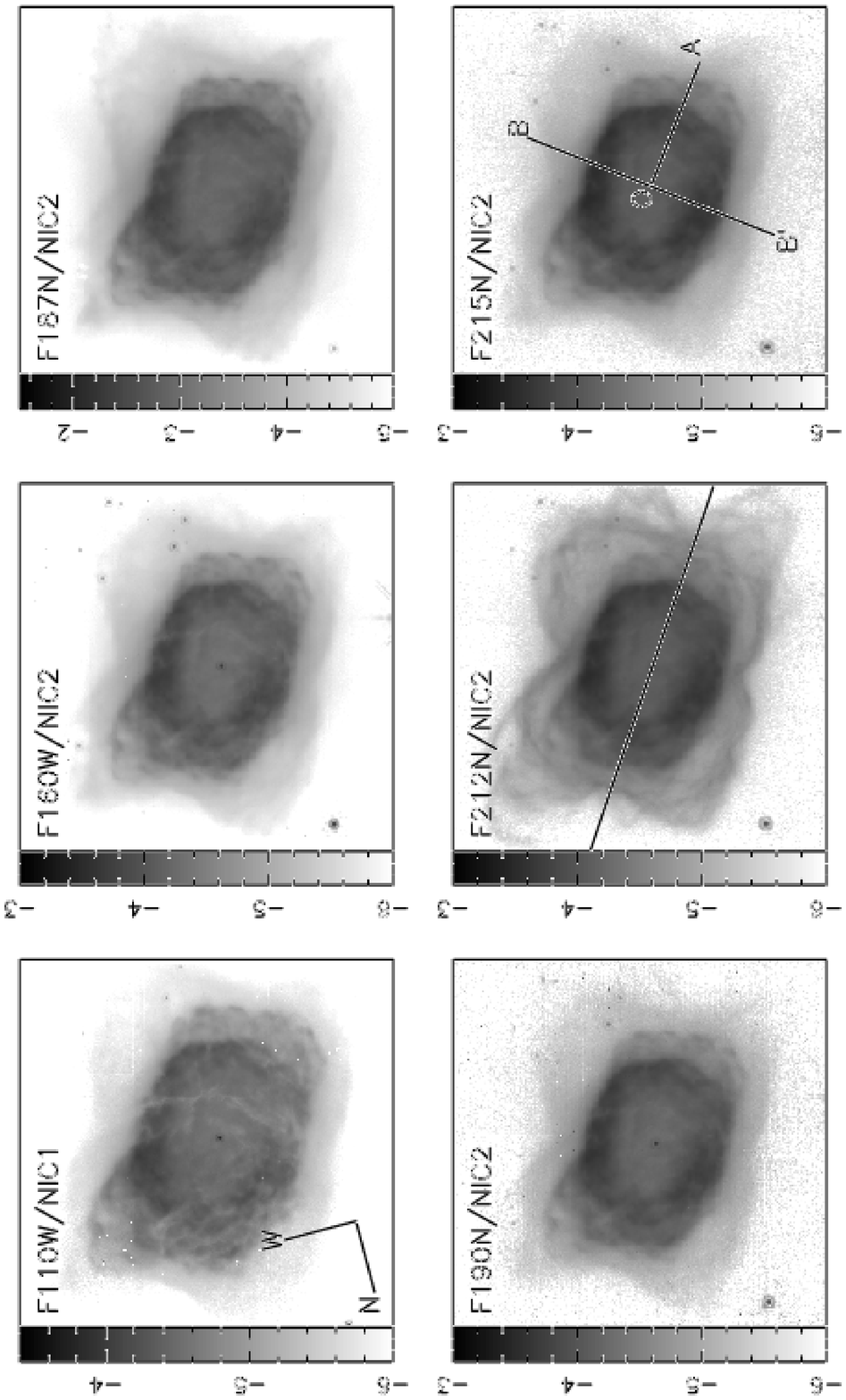}
\figcaption[fig1.eps]{
NICMOS/HST mosaiced images of NGC 7027 covering 1.10 $\mu$m~to 2.15 $\mu$m.
The F110W/NIC1 image has a pixel scale of 0.043\arcsec~pixel$^{-1}$ and 
a field of view of 16\arcsec$\times$16\arcsec; the other images have a
scale of 0.075\arcsec~pixel$^{-1}$ and a field of view of 19.2\arcsec$\times$19.2\arcsec. 
The images are in units of Jy pixel$^{-1}$, and are displayed here using 
a logarithmic stretch. The orientation of the CSHELL slit is shown in 1(e) and
1(f) shows directions for the radial cuts in Fig. 7. 
\label{Figure 1}}
\end{figure}

\begin{figure}
\epsscale{0.5}
\plotone{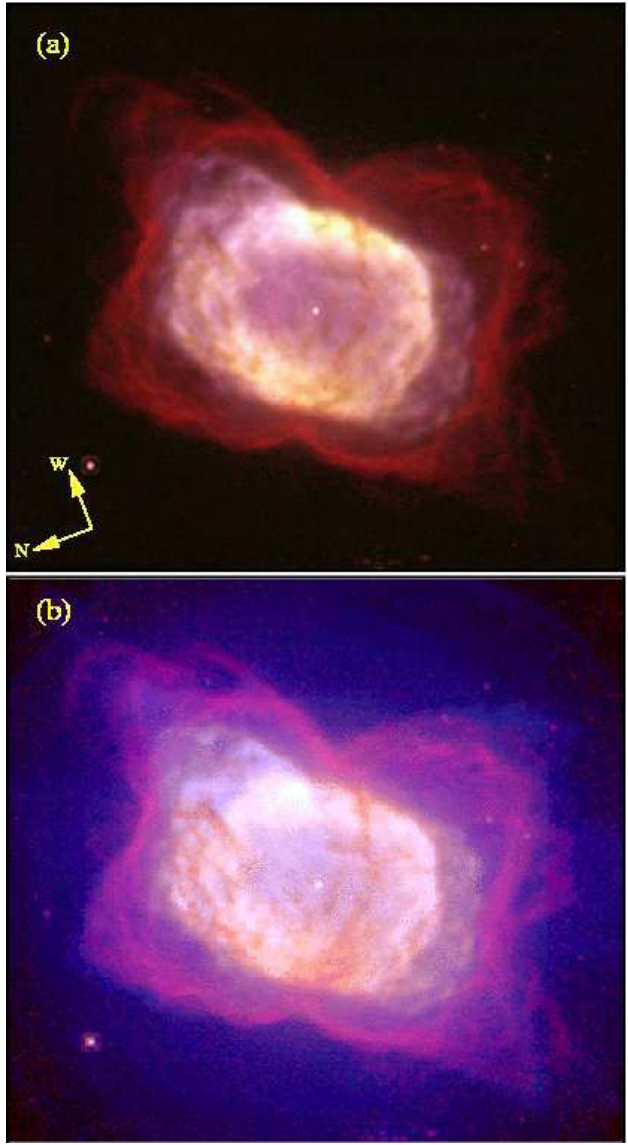}
\figcaption[fig2.ps]{{\it (a)}
Three--color composite of data taken in three NICMOS
filters. Red is F212N (molecular hydrogen), green is F215N (continuum),
and blue is F110W (continuum). These data have been highly stretched for
visualization and the relative brightness between bands has not been
maintained. It is evident from this image that
the dark ``ring'' going across the bright core, is glowing in $\lambda$=2.121 
\micron~molecular hydrogen emission (seen in red/orange color) that is
contiguous with the more extended emission. When compared with the
individual filter images (Fig. 1), it is clear
that there is a strong correlation between the \h2\ emission and
regions of high attenuation.
{\it (b)} Three-color composite of two NICMOS images and one acquired
with WFPC2. Red is NICMOS F212N, green is NICMOS F190N (continuum), and
blue is WFPC2 V and I bands combined. As with the images in {\it (a)}
(NICMOS), the combined data has been highly stretched for
visualization.  Many of the same features seen in {\it (a)} are visible
here as well. Dust scattered visible light is seen in these data as the
extended blue emission.
\label{colorimages}}
\end{figure}

\begin{figure}
\epsscale{0.8}
\plotone{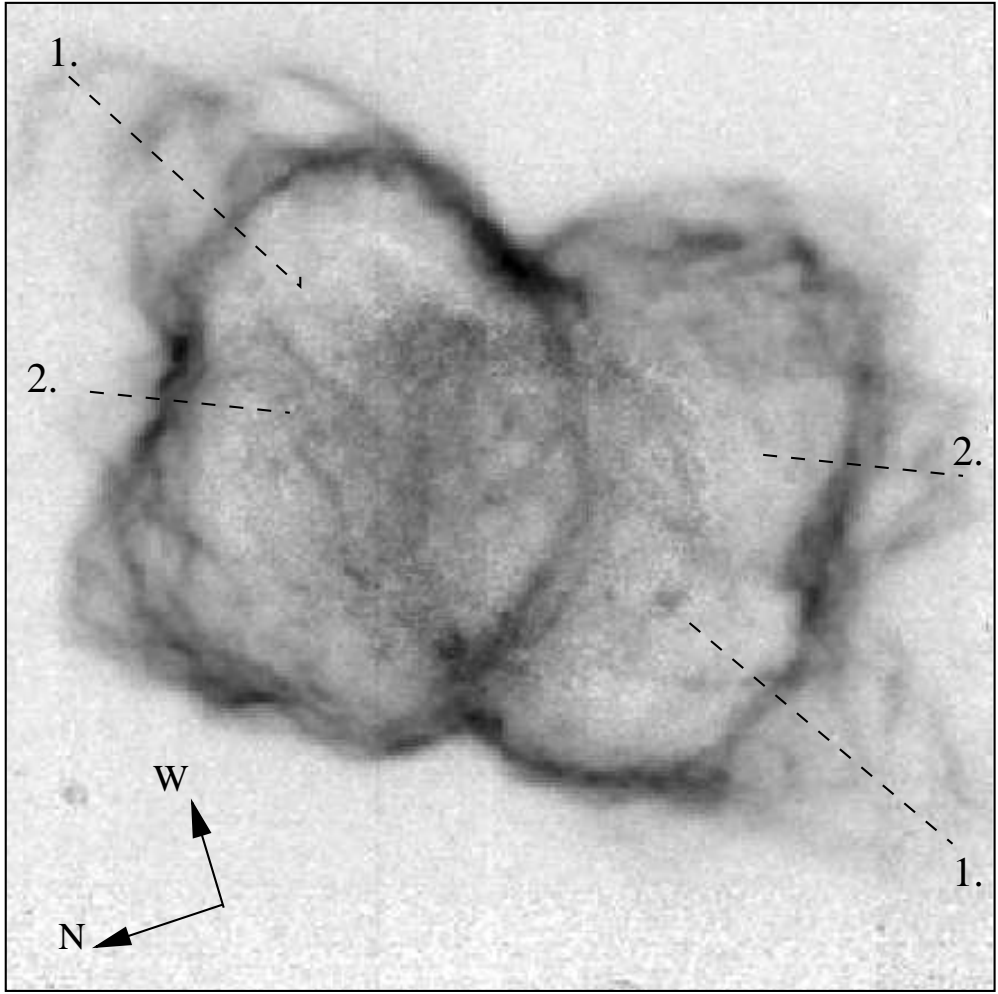}
\figcaption[fig3.ps]{Continuum-subtracted H$_2$ image. The two
intersecting rings are interpreted as the edges of a tilted bicone
(see text). Two outflows/jets, labelled 1 and 2, appear to have
disturbed the general morphology. Field of View is $\sim$ 20\arcsec~
$\times$ 20\arcsec.
\label{h2only}}
\end{figure}

\begin{figure}
\epsscale{0.8}
\plotone{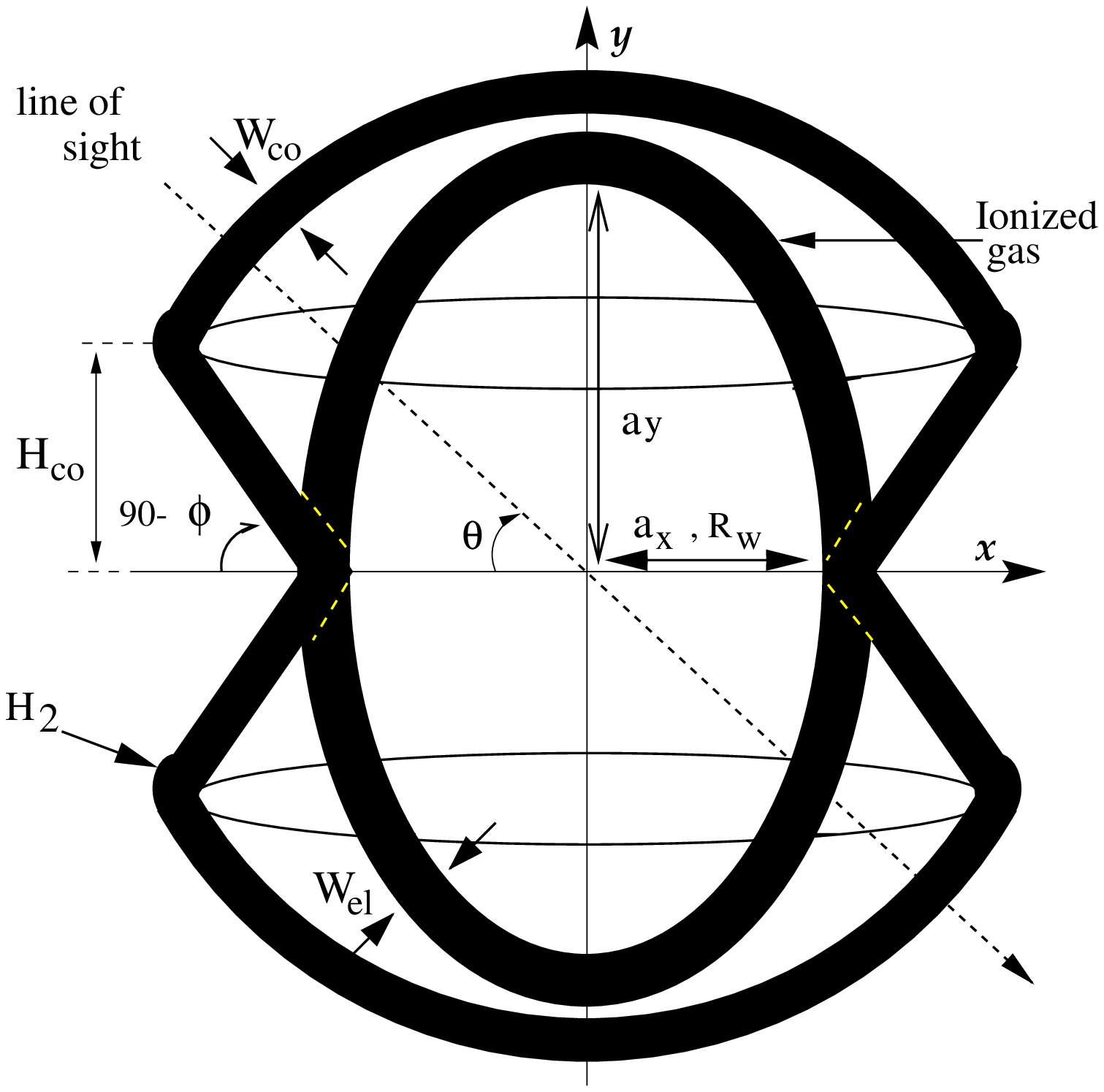}
\figcaption[fig4.ps]{A schematic of the 2--component
axially symmetric model. The bicone and the ellipsoid share the same
inner radius in the equatorial (X-Z) plane, i.e. a$_x$ = R$_w$ (see text
for details). The widths of both components, W$_{co}$, W$_{el}$, 
are derived independently but assumed to be constant. $\theta$ is the tilt
of the nebula to the line of sight, $\phi$ is the semi-opening angle of the
bicone, H$_{co}$ is the semi-height of the cone and a$_x$, a$_y$ and a$_z$ 
are the semi-axes of the ellipsoid. Both of the ``caps" are sections of a 
sphere centered at the origin.
\label{modelschem}}
\end{figure}

\begin{figure}
\epsscale{1.0}
\plotone{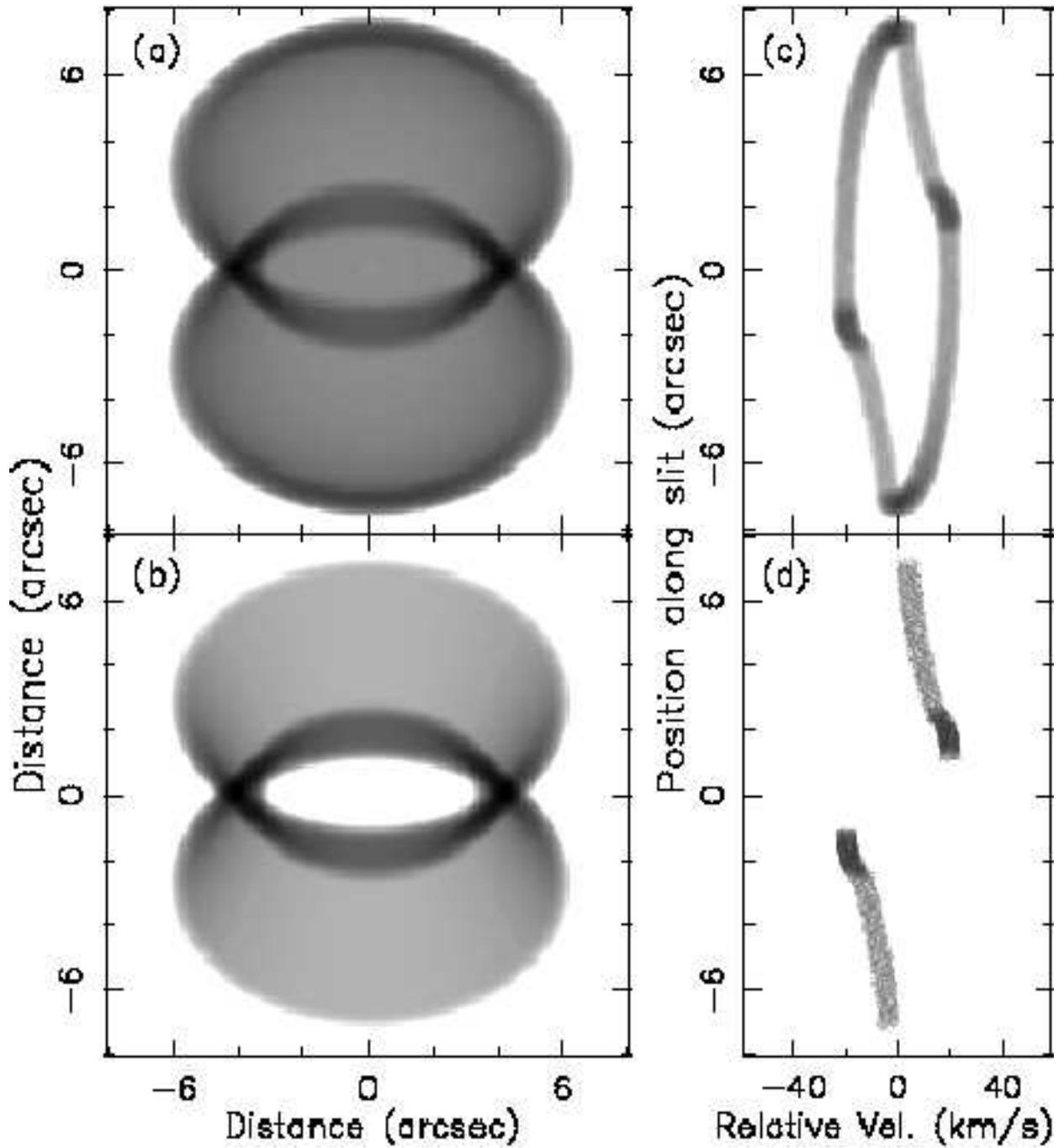}
\figcaption[fig5.ps]{Bicone model images and Position--Velocity diagrams obtained by
placing a slit along the long axis of the nebula.
{\it (a), (c)}: ``Best-fit" model bicone with ``cap"; {\it (b), (d)}: bicone without
``cap". 
\label{capdig}}
\end{figure}

\begin{figure}
\epsscale{0.65}
\plotone{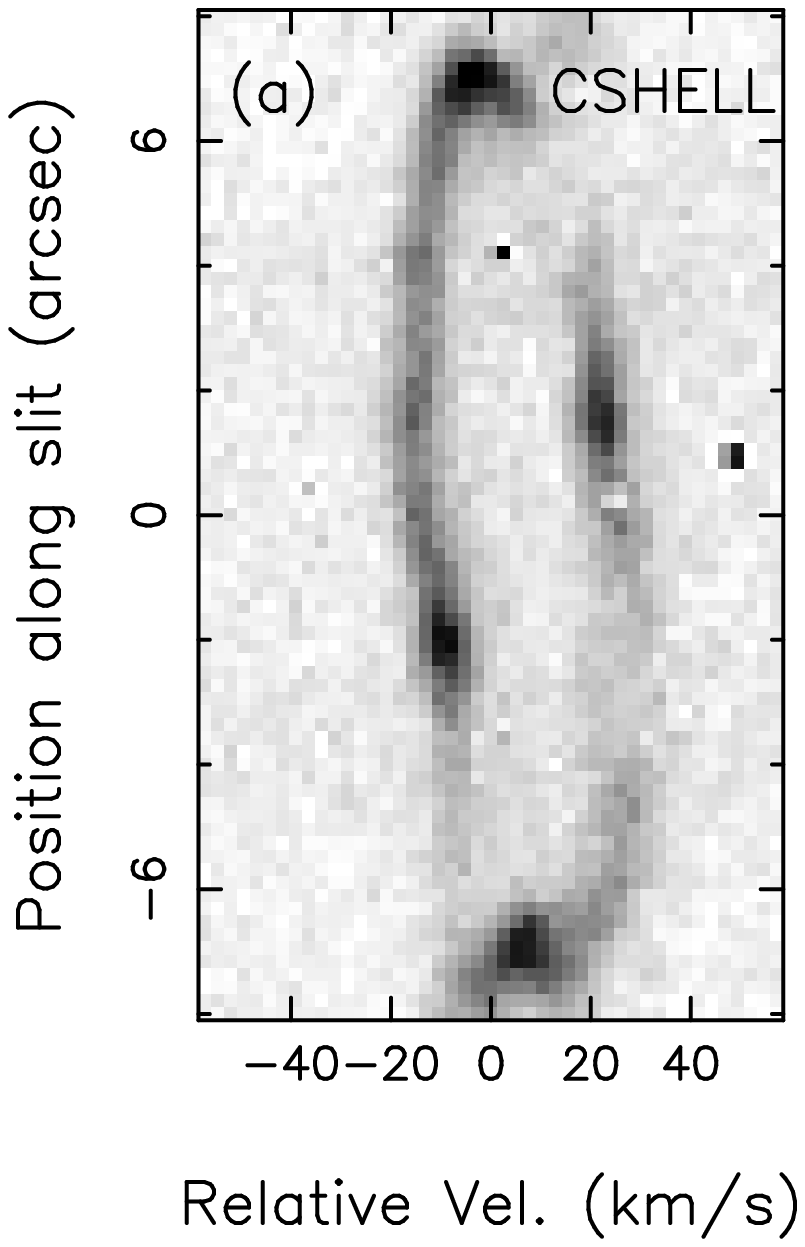}
\end{figure}
\begin{figure}
\epsscale{0.5}
\plotone{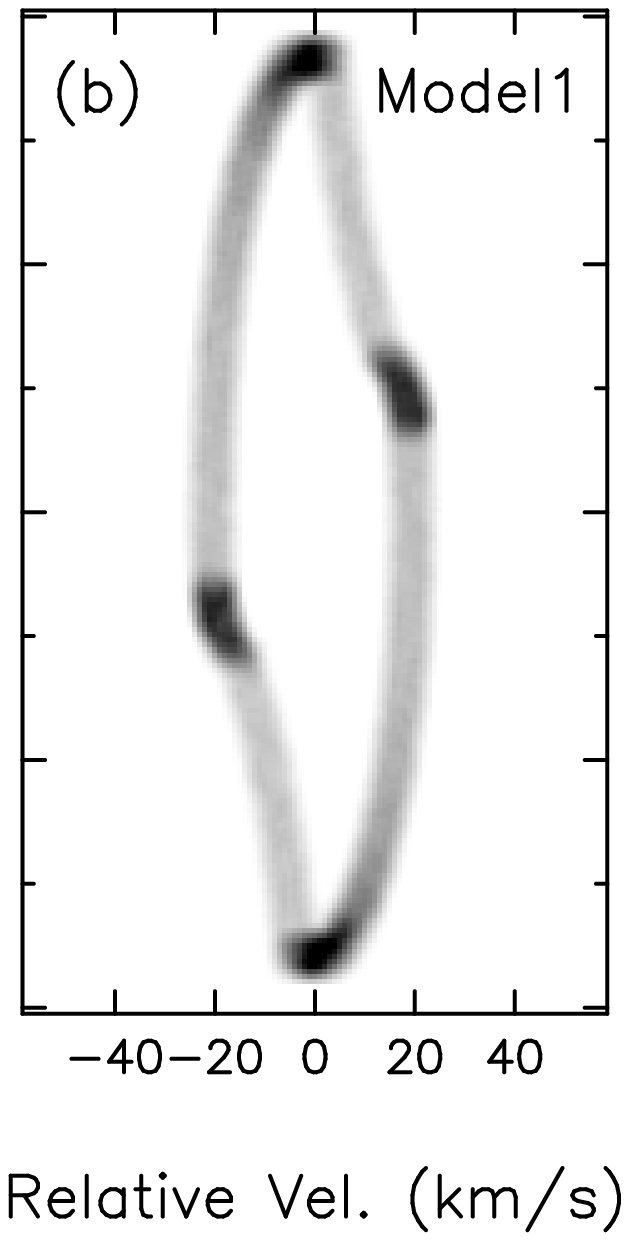}
\end{figure}
\begin{figure}
\epsscale{0.5}
\plotone{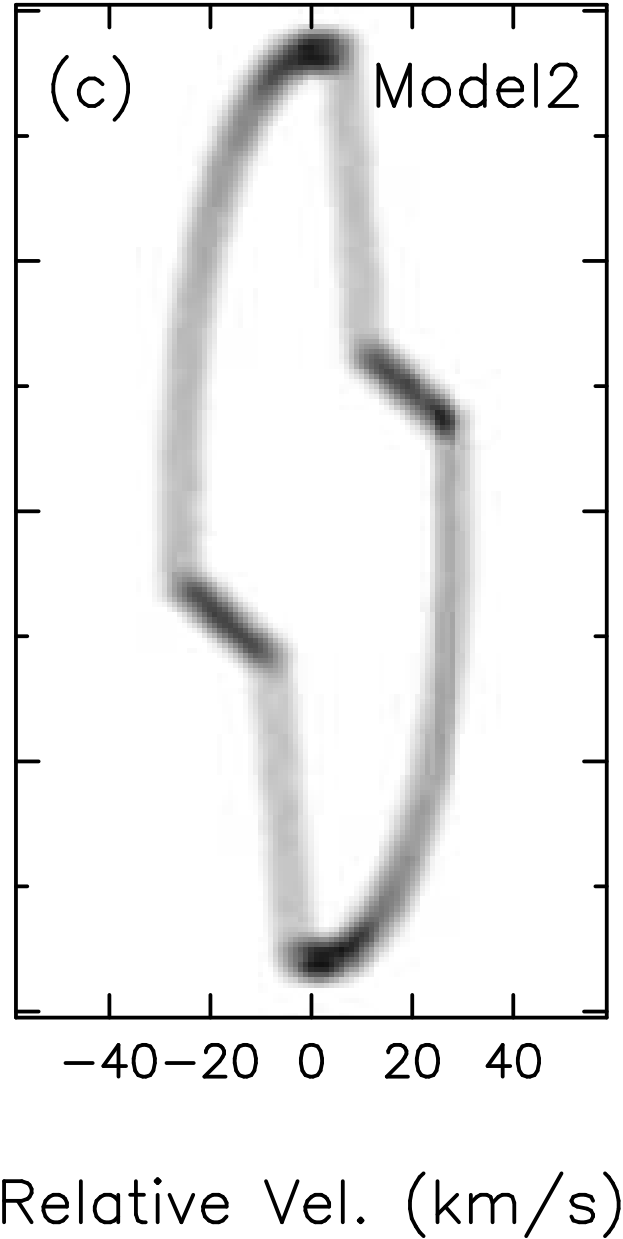}
\figcaption[fig6a.ps]{Observed Position--Velocity diagram obtained by
placing the CSHELL slit along the major axis of NGC 7027 (see Fig. 1(e)), compared
with model P-V diagrams, shown on the same spatial/velocity scale.
{\it (a)} The CSHELL observations have a velocity scale of 2.6\arcsec~
pixel$^{-1}$. {\it (b)} Best-fit model
with a constant velocity, V = 20 km s$^{-1}$. {\it (c)} Model with
varying velocity, V $\propto$ R$^{1.0}$ km s$^{-1}$.
\label{posveldig}}
\end{figure}

\begin{figure}
\epsscale{0.8}
\plotone{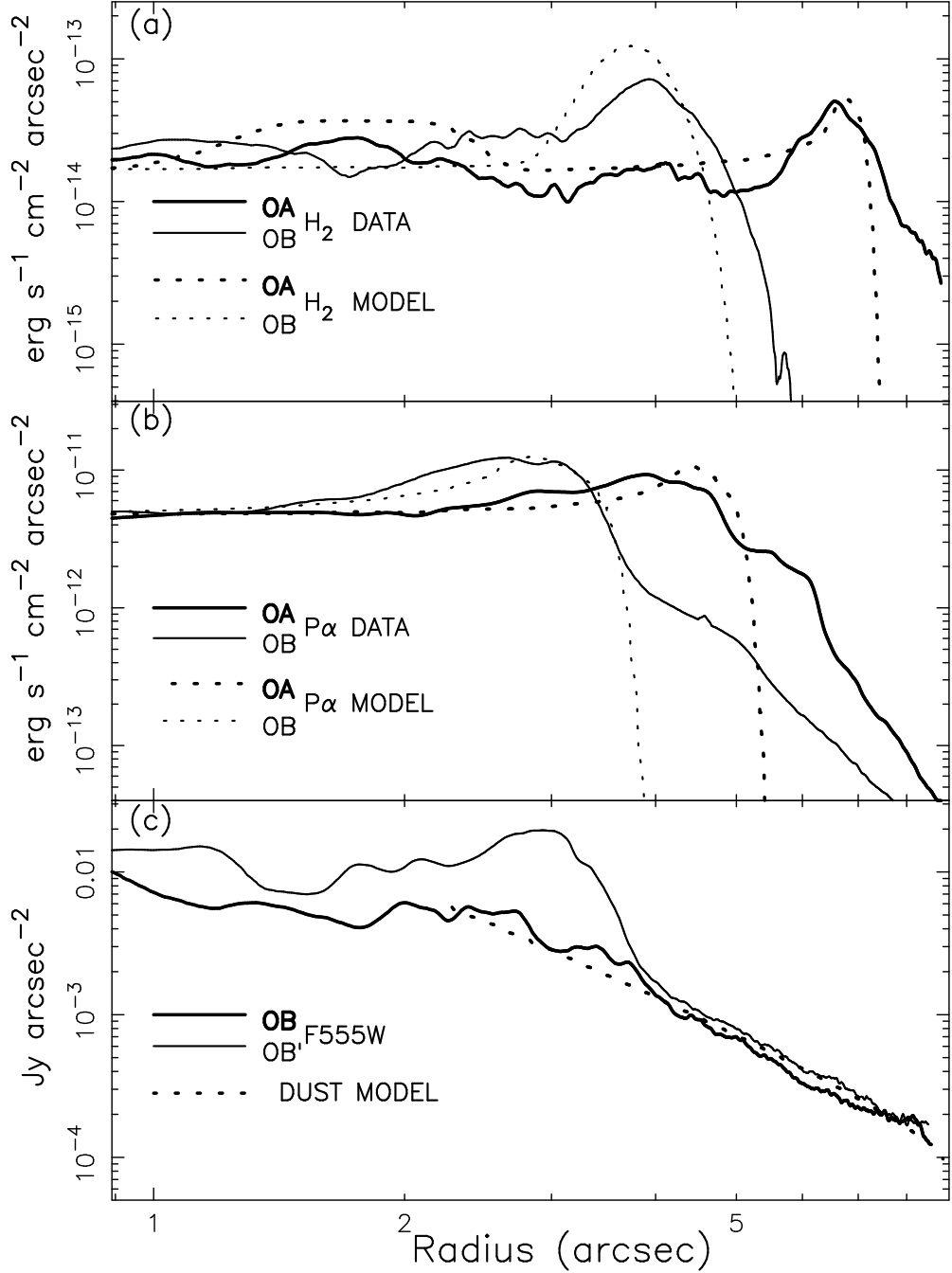}
\figcaption[fig7.ps]{Radial cuts across images. The direction of the
cuts (OA, OB, OA') are shown in Fig. 1, F215N panel.  {\it (a)}:
Radial cuts across the continuum subtracted H$_2$ image and the H$_2$
bicone model.  Dark solid line is a cut along the major axis, faint
solid line is a cut along the minor axis. The dotted lines are
corresponding radial cuts through the bicone model.  {\it (b)}: Radial
cuts across the continuum subtracted \Pa\ image and the constant--density ellipsoid
model.  Dark solid line is a cut along the major axis, faint solid
line is a cut along the minor axis. The dotted lines are corresponding
radial cuts through the ellipsoid model.  {\it (c)}: Radial cuts along
the minor axis of the F555W image (solid lines) overlayed on a radial
cut from the optically thin, uniform, spherical model of scattering
dust grains.
\label{radialcuts}}
\end{figure}

\begin{figure}
\epsscale{0.7}
\plotone{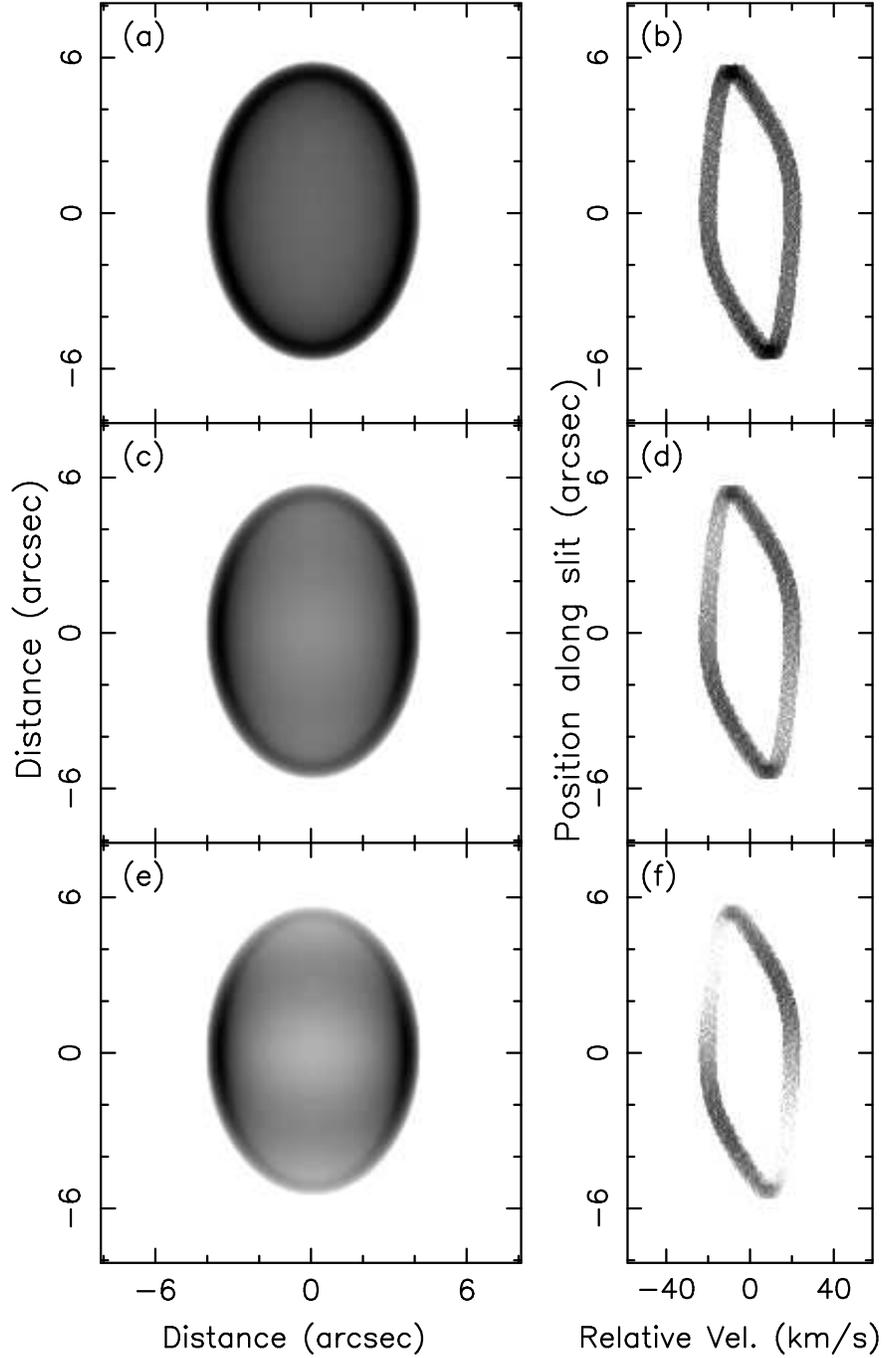}
\figcaption[fig8.ps]{Ellipsoidal shell models -- Images and Position--Velocity
diagrams for the distribution of ionized gas. In each model the velocity is
assumed to be constant. {\it (a), (b)}: Constant
density; {\it (c), (d)}: Density $\propto$ R$^{-1}$; {\it (e), (f)}:
Density $\propto$ R$^{-2}$. The equator--to--pole intensity contrasts
in the models with decreasing density are much higher than the
observed intensity contrast, suggesting that density changes by
$\lesssim$ 20\% -- 30\% going from the equator to pole.
\label{ellip-panel}}
\end{figure}

\newpage
\begin{deluxetable}{lcccc}
\tablecolumns{5}
\tablecaption{Summary of Observations\label{tblsumobs}}
\tablehead{
\colhead{Camera}  & \colhead{Filter}   &  \colhead{$\lambda_{eff}$ ($\mu$m)} &
\colhead{$\Delta\lambda_{eff}$ ($\mu$m)}        &  \colhead{Integration Time (s)}
}
\startdata
  NIC1  	  & 110W     & 1.1022   & 0.5920     &  256 \nl
  NIC2            & 160W     & 1.5931   & 0.4030    &  256 \nl
  NIC2            & 187N     & 1.8738   & 0.0192     &  1024 \nl
  NIC2            & 190N     & 1.9003   & 0.0174    &  1024 \nl
  NIC2            & 205W     & 2.0406   & 0.6125     &  1024 \nl
  NIC2            & 212N     & 2.121   & 0.021     &  2560 \nl
  NIC2            & 215N     & 2.149   & 0.020     &  2176 \nl
\enddata
\end{deluxetable}

\begin{deluxetable}{cccccccc}
\tablecolumns{8}
\tablecaption{Model Parameters for NGC 7027\label{tblmodpar}}
\tablehead{
\colhead{Component}     &
\colhead{Shape} & \colhead{Tilt}  & \colhead{}  &
\multicolumn{4}{c}{Parameters\tablenotemark{a}}\\
}
\startdata
H$_2$       & Bicone   & 45$^{0}$ & & $\phi$\tablenotemark{b} & R$_w$\tablenotemark{b}  & H$_{co}$\tablenotemark{b}  & Width (W$_{co}$) \nl
            &          &          & & 30$^0$  & 2.8\arcsec    & 3.9\arcsec & 
0.5\arcsec   \nl  
Ionized Gas & Ellipsoid& 45$^{0}$ & & a$_x$   & a$_y$  & a$_z$ & Width (W$_{el}$) \nl
            &          &          & & 2.8\arcsec & 5.3\arcsec  & 2.8\arcsec & 0.9\arcsec \nl
Cool dust   & Sphere   &   --     & & R$_{in}$ & R$_{out}$ & &  \nl
	    &          &          & & 3.0\arcsec    & 8.0\arcsec    &   &  
\enddata
\tablenotetext{a}{Dimensions given in arcsec. At the adopted distance
of 880 pc, 1\arcsec=1.32$\times$10$^{16}$ cm.}
\tablenotetext{b}{$\phi$ is the semi opening--angle of the cone, R$_w$ is
the radius of the waist of the cone and H$_{co}$ is the semi-height of the cone.}
\end{deluxetable}

\begin{deluxetable}{lccccccc}
\tablecolumns{7}
\tablewidth{7.0in}
\tablecaption{Photometry and Zanstra Temperature}
\tablehead{
\colhead{Filter} & \colhead{Description} & 
\multicolumn{2}{c}{Stellar Photometry} &
\multicolumn{2}{c}{Nebular Photometry} & \colhead{$T_{Z}$} \\ 
\cline{3-4} \cline{5-6}
\colhead{}  &  \colhead{}  &  \colhead{F$_{\nu,obs}$} &
\colhead{F$_{\nu,0}$\tablenotemark{a}} &
\colhead{F$_{\nu,obs}$} & \colhead{F$_{\nu,0}$\tablenotemark{a}} &\\
\colhead{}  &  \colhead{}  &  \colhead{(mJy)} &
\colhead{(mJy)} &
\colhead{(mJy)} & \colhead{(mJy)} &
\colhead{(K)}
}
\startdata
F205W &	continuum & 1.197$\pm$0.06 & 1.68$\pm$0.10 &   2.776$\pm$0.14 & 3.739$\pm$0.40 & 192,000$\pm$7,700 \nl
F190N &	P$\alpha$ cont. & 1.308$\pm$0.07 & 1.88$\pm$0.11 &  1.36$\pm$0.01 & $1.91\pm0.08$ & 190,000$\pm$8,600 \nl
F160W &	continuum & 1.377$\pm$0.07 & 2.30$\pm$0.18 &   1.082$\pm$0.05 & 1.693$\pm$0.19 & 207,000$\pm$8,300 \nl
F110W &	continuum & 1.765$\pm$0.09 & 5.98$\pm$0.57 &   1.394$\pm$0.07 & 3.671$\pm$0.56 & 198,500$\pm$8,900 \nl
F814W\tablenotemark{b} &  continuum  & 1.362$\pm$0.14 & 7.78$\pm$1.44 &  
0.875$\pm$0.09 & 4.464$\pm$0.60  & 214,000$\pm$18,200 \nl 
F555W\tablenotemark{c} &  continuum  & 0.977$\pm$0.10 & 25.03$\pm$6.38 &  
0.922$\pm$0.09 & 14.214$\pm$2.27  & 187,300$\pm$20,600 \nl 

F187N	& P$\alpha$ line+cont. & --------- & --------- &   24.64$\pm$0.62 &  
$34.81\pm1.39$	& ----------- \nl
F212N\tablenotemark{d}	& H$_2$ line+cont.  & --------- & --------- &  
1.92$\pm$0.05& $2.53\pm0.09$ & ----------- \nl
F215N\tablenotemark{d}	& H$_2$ cont. & --------- & --------- &   1.81$\pm$0.05 & $2.36\pm0.08$  & ----------- \nl

\enddata 
\tablenotetext{a} {The flux density has been corrected for
attenuation using the mean extinction curve in Table 3.1 of Whittet
(1992), and A$_V$ = 3.17 mag for the CS and 2.97 mag for mean nebular
attenuation (Robberto et al. 1993).} 
 \tablenotetext{b} {Images taken
by the WFPC2/HST, extracted from the HST archive.  $\lambda_{eff}$ =
0.811$\mu$m; $\Delta\lambda$ = 0.176$\mu$m} 
\tablenotetext{c} {Images
taken by the WFPC2/HST, extracted from the HST archive.
$\lambda_{eff}$ = 0.513$\mu$m; $\Delta\lambda$ = 0.122$\mu$m}
\tablenotetext{d} {The F212N filter is contaminated by He I and the
F215N filter by Br$\gamma$. See \S 3.1.1 for a discussion.}
\end{deluxetable}

\begin{deluxetable}{lc}
\tablenum{4}
\tablecolumns{2}
\tablewidth{6.5in}
\tablecaption{Derived Properties of the Central Star}
\tablehead{
\colhead{Stellar Property} & \colhead{Value} }
\startdata
Zanstra temperature & 198,100 $\pm$ 10,500 K \nl
Luminosity\tablenotemark{a} & 7,710 $\pm$ 860 L$_{\odot}$ \nl
Radius & 5.21 $\times 10^9$ cm \nl
Core mass\tablenotemark{b} & 0.7 M$_{\odot}$ \nl
Initial mass\tablenotemark{b} &  4 M$_{\odot}$ \nl
Evolutionary timescale & 700 y \nl
\enddata
\tablenotetext{a}{Based on best fit stellar temperature and a distance of 880 pc.}
\tablenotetext{b}{From evolutionary tracks of Bl\"{o}cker (1995).}

\end{deluxetable}

\end{document}